\documentclass[a4paper,10pt,eqno]{article}
\usepackage{theorem}
\usepackage{latexsym,amssymb,amsfonts,amsmath}
\usepackage{graphicx}
\theoremheaderfont{\scshape}
\newcommand{\RM}{\mathbb{R}}
\newcommand{\ZM}{\mathbb{Z}}

\newcommand{\CM}{\mathbb{C}}

\newcommand{\ket}[1]{|#1 \rangle}

\title{{\Large {\bf A new time-series model based on quantum walk}}}
\author{
{\small Norio Konno}\\
{\scriptsize Department of Applied Mathematics, 
Faculty of Engineering, 
Yokohama National University}\\
{\scriptsize Hodogaya, Yokohama 240-8501, Japan}\\
{\scriptsize e-mail: konno@ynu.ac.jp}\\
}
\date{\empty }
\begin{document}
\maketitle

\par\noindent
\begin{small}
\par\noindent
{\bf Abstract}. The quantum walk (QW) was introduced as a quantum counterpart of the classical random walk. A number of non-classical properties of the QW have been shown, e.g., ballistic spreading, anti-bellshaped limit density, localization. Since around 2000, extensive research has been conducted in both theoretical aspects as well as the practical application of QWs. However, the application of a QW to the time-series analysis is not known. On the other hand, it is well known that the ARMA or GARCH models have been widely used in economics and finance. These models are studied under some suitable stationarity conditions. In this paper, we propose a new time-series model based on the QW, which does not assume such a stationarity. Therefore, our method would be applicable to the non-stationary time series.

\footnote[0]{
{\it Abbr. title:} A new time-series model based on quantum walk
}
\footnote[0]{
{\it AMS 2000 subject classifications: }
60F05, 60G50, 82B41, 81Q99
}
\footnote[0]{
{\it PACS: } 
03.67.Lx, 05.40.Fb, 02.50.Cw
}
\footnote[0]{
{\it Keywords: } 
Quantum walk, time-series model, stationarity
}
\end{small}

\setcounter{equation}{0}
\section{Introduction}
The quantum walk (QW) was introduced as a quantum counterpart of the classical random walk and it has extensively studied since around 2000 (\cite{Gudder1988, AharonovEtAl1993, Meyer1996, AmbainisEtAl2001}). A number of non-classical properties of the QW have been shown, for example, ballistic spreading, anti-bellshaped limit density, and localization. Currently, active research is being conducted in both theoretical aspects as well as the practical application of QWs. The various proposals regarding QW application methods include the strongly correlated electron system \cite{OkaEtAl2005}, topological insulators \cite{Kitagawa2012,OK2011}, and radioactive waste reduction \cite{IMKY2013,MKHY2011,MY2013}. Several books and reviews related to QWs have been published, for instance, Venegas-Andraca \cite{VAndraca2008, Venegas2013}, Konno \cite{Konno2008b}, Cantero et al. \cite{CanteroEtAl2012b}, Manouchehri and Wang \cite{MW2013}, Portugal \cite{P2013}. However, the application of a QW to the time-series analysis is not known. On the other hand, it is well known that the ARMA or GARCH models have been widely used in economics and finance, see Box and Jenkins \cite{BoxJenkins1976}, Engle \cite{Engle1982}, Bollerslev \cite{Bollerslev1986}. One usually considers the ARMA or GARCH models under some suitable stationarity conditions. In this paper, we present a new time-series model based on the QW, which does not assume such a stationarity. Therefore, our method would be applicable to the non-stationary time-series. As such a candidate, time series like a cryptocurrency such as Bitcoin can be considered. Concerning Bitcoin, see Nakamoto's very famous paper as founder \cite{Nakamoto2008}. Of course it would also be applicable to time series other than economy and finance. We should remark that our method is not related to an It\^o's formula for the discrete-time QW proposed by the author \cite{Konno2013}. This paper is the first step for a time-series analysis via the QW.

The rest of the manuscript is organized as follows. Section \ref{def}  is devoted to the definition of the QW considered here. In Section \ref{new}, we propose a new time-series model using the QW. Section \ref{1d2s} deals with a two-state one-dimensional QW case. Finally, Section \ref{con} concludes the paper.

\section{Definition of QW \label{def}}
Let us first give a definition of the $m$-state QW on a $d$-dimensional integer lattice, $\ZM^d$, where $\ZM$ is the set of integers. The Hilbert space of the quantum walker is given by the tensor product
\begin{align*}
{\cal H} = {\cal H}_p \otimes {\cal H}_c,
\end{align*}
where ${\cal H}_p$ is the position space defined by 
\begin{align*}
{\cal H}_p = {\rm Span} \{ | {\bf x} \rangle \> ; \> {\bf x} \in \ZM^d \}.
\end{align*}
Moreover, ${\cal H}_c$ is the coin space with an $m$-state given by
\begin{align*}
{\cal H}_c = {\rm Span} \{ | C_1 \rangle, \> | C_2 \rangle, \ldots, | C_m \rangle \}.
\end{align*}
Here $\{ | {\bf x} \rangle \> ; \> {\bf x} \in \ZM^d \}$ and $\{ | C_1 \rangle, \> | C_2 \rangle, \ldots, | C_m \rangle \}$ are complete orthonormal bases of ${\cal H}_p$ and  ${\cal H}_c$, respectively. For example, in the case of $m=2$ and $d=1$ (two-state one-dimensional QW), the quantum walker on a line, $\ZM$, has two possibilities (or chiralities), $| L \rangle = | C_1 \rangle$ (left) and $| R \rangle = | C_2 \rangle$ (right), in each step - it can move to the left or right. To each of these options, we assign a vector of the standard basis of the coin space 
\begin{align*}
{\cal H}_c = \CM^2 = {\rm Span} \{ | L \rangle, \> | R \rangle \},
\end{align*}
such as 
\begin{align*}
\ket{L} = 
\begin{bmatrix}
1 \\
0
\end{bmatrix}
,
\qquad
\ket{R} = 
\begin{bmatrix}
0 \\
1
\end{bmatrix}
.
\end{align*}
Here $\CM$ is the set of complex numbers. In our setting, $| L \rangle$ corresponds to the fact that the value in time series decreases by 1. Similarly, $| R \rangle$ corresponds to the fact that the value in time series increases by 1. A single step of the QW is determined by the unitary operator $U^{(s)}$ (superscript $s$ means the initial of {\it system}) given by
\begin{align*}
U^{(s)} = S \cdot \left( I_p \otimes U \right),
\end{align*}
where $S$ is the shift operator, $I_p$ denotes the identity on the position space ${\cal H}_p$ and $U$ is a unitary operator called {\it quantum coin} on the coin space ${\cal H}_c$. If a quantum walker starts from the origin with a coin state $| C_1 \rangle$, then the initial state of the QW is given by $| {\bf 0} \rangle \otimes | C_1 \rangle \in {\cal H}_p \otimes {\cal H}_c.$ For instance, in the case of $m=2$ and $d=1$ (two-state one-dimensional QW), $S$ has the following form
\begin{align*}
S = \sum_{x \in \ZM} \left( |x-1 \rangle \langle x | \otimes | L \rangle \langle L | + |x+1 \rangle \langle x | \otimes | R \rangle \langle R | \right).
\end{align*}
If we choose the Hadamard matrix as the quantum coin, 
\begin{align*}
U = 
\frac{1}{\sqrt{2}}
\begin{bmatrix}
1 & 1 \\
1 & -1
\end{bmatrix}
,
\end{align*}
then the QW is the well-known {\it Hadamard walk} which has been intensively investigated in the study of the QW. Another equivalent definition of the QW based on the path counting will be presented in Section \ref{1d2s}.

Define parameter sets $\Theta_1 \subset \RM^{M_1}$ and $\Theta_2 \subset \RM^{M_2}$, where $M_1, M_2 \in \ZM_{>}$, where $\RM$ is the set of real numbers and $\ZM_{>} = \{1,2, \ldots \}$. For each parameter $\theta_k = (\theta^{(1)}_k, \theta^{(2)}_k, \ldots, \theta^{(M_k)}_k) \in \Theta_k \> (k=1,2)$, we consider the corresponding quantum coin $U = U(\theta_1) = U (\theta^{(1)}_1, \theta^{(2)}_1, \ldots, \theta^{(M_1)}_1)$ and the corresponding initial state $\varphi = \varphi (\theta_2) = \varphi (\theta^{(1)}_2, \theta^{(2)}_2, \ldots, \theta^{(M_2)}_2)$ at the origin.

In Section \ref{1d2s}, we will deal with a time-series analysis based on a set of two-state one-dimensional QWs with the following setting. Put $\Theta_1=[0,1], \> \Theta_2=[0, \pi/2] \> (M_1=M_2=1)$ and $\theta_1 = \theta^{(1)}_1, \theta_2 = \theta^{(1)}_2$. Moreover, we define a one-parameter set of quantum coins:
\begin{align}
U = U \left( \theta^{(1)}_1 \right) = 
\begin{bmatrix}
\theta^{(1)}_1 & \sqrt{1- (\theta^{(1)}_1)^2} \\
\sqrt{1- (\theta^{(1)}_1)^2} & - \theta^{(1)}_1
\end{bmatrix},
\label{funfair01}
\end{align}
where $\theta^{(1)}_1 \in \Theta_1=[0, 1]$. We should note that the QW determined by the coin parameter $\theta^{(1)}_1 = 1/\sqrt{2}$ becomes the Hadamard walk. Moreover, we define a one-parameter set of initial states:
\begin{align}
\varphi = \varphi \left( \theta^{(1)}_2 \right) =
\begin{bmatrix}
\cos \theta^{(1)}_2 \\
i \> \sin \theta^{(1)}_2
\end{bmatrix}
.
\label{funfair02}
\end{align}
Here $\theta^{(1)}_2 \in \Theta_2 = [0, \pi/2].$ From Theorem 4 in Konno \cite{Konno2002}, we see that a necessary and sufficient condition that the probability distribution of a QW with parameter $(\theta^{(1)}_1, \theta^{(1)}_2)$ for any time $n \in \ZM_{\ge}$ is symmetric with respect to the origin is ``$\theta^{(1)}_2 = \pi/4$", where $\ZM_{\ge} = \{0,1,2, \ldots \}$.

If we treat a time-series $\{ x_0, x_1, x_2, \ldots \}$ in which some values do not change, e.g., $x_n = x_{n+1}$, we take a time-series analysis using a set of three-state one-dimensional QWs. Then we consider the following setting whose quantum coin was introduced by Stefanak et al. \cite{StefanakEtAl2012} as such a candidate. The position space is 
\begin{align*}
{\cal H}_p = {\rm Span} \{ | {\bf x} \rangle \> ; \> {\bf x} \in \ZM \}
\end{align*}
and the coin space is 
\begin{align*}
{\cal H}_c = \CM^3 = {\rm Span} \{ | L \rangle, \> | S \rangle, \> | R \rangle \},
\end{align*}
such as 
\begin{align*}
\ket{L} = 
\begin{bmatrix}
1 \\
0 \\
0
\end{bmatrix}
,
\qquad
\ket{S} = 
\begin{bmatrix}
0 \\
1 \\
0
\end{bmatrix}
,
\qquad
\ket{R} = 
\begin{bmatrix}
0 \\
0 \\
1
\end{bmatrix}
.
\end{align*}
The quantum walker on a line, $\ZM$, can move to the left $\ket{L}$ or right $\ket{R}$ or stay at its current position $\ket{S}$ in each step. Then $\ket{S}$ corresponds to the fact that the value in time series does not change.

Put $\Theta_1=[0,1], \> \Theta_2=[0, 2 \pi]^2 \> (M_1=1, \> M_2=2)$ and $\theta_1 = \theta^{(1)}_1, \theta_2 = (\theta^{(1)}_2, \theta^{(2)}_2)$. We introduce a one-parameter set of quantum coins: 
\begin{align}
U = U \left( \theta^{(1)}_1 \right) = 
\begin{bmatrix}
- (\theta^{(1)}_1)^2 & \theta^{(1)}_1 \sqrt{2 \left( 1- (\theta^{(1)}_1)^2 \right)}  & 1 - (\theta^{(1)}_1)^2 \\
\theta^{(1)}_1 \sqrt{2 \left( 1- (\theta^{(1)}_1)^2 \right)} & 2 (\theta^{(1)}_1)^2 - 1 & \theta^{(1)}_1 \sqrt{2 \left( 1- (\theta^{(1)}_1)^2 \right)} \\
1 - (\theta^{(1)}_1)^2 & \theta^{(1)}_1 \sqrt{2 \left( 1- (\theta^{(1)}_1)^2 \right)}  & - (\theta^{(1)}_1)^2
\end{bmatrix},
\label{funfair03}
\end{align}
where $\theta^{(1)}_1 \in \Theta_1=[0, 1]$. Remark that the QW determined by the coin parameter $\theta^{(1)}_1 = 1/\sqrt{3}$ becomes the three-state Grover walk on $\ZM$. Moreover, we define a two-parameter set of initial states: 
\begin{align}
\varphi = \varphi \left( \theta^{(1)}_2, \theta^{(2)}_2 \right) =
\frac{1}{\sqrt{3}}
\begin{bmatrix}
1 \\
\exp (i \> \theta^{(1)}_2) \\
\exp (i \> \theta^{(2)}_2)   
\end{bmatrix}
,
\label{funfair04}
\end{align}
where $(\theta^{(1)}_2, \theta^{(2)}_2) \in \Theta_2 = [0, 2 \pi]^2.$

We will move to another example. For example, in order to estimate the price of a cryptocurrency, we would need to consider not only price but also volume of the cryptocurrency in which both values have a correlation. So we treat such a data by using a set of four-state two-dimensional QWs with the following setting. The position space is 
\begin{align*}
{\cal H}_p = {\rm Span} \{ | {\bf x} \rangle \> ; \> {\bf x} \in \ZM^2 \}
\end{align*}
and the coin space is 
\begin{align*}
{\cal H}_c = \CM^4 = {\rm Span} \{ | L \rangle, \> | R \rangle, \> | D \rangle  , \> | U \rangle \},
\end{align*}
such as 
\begin{align*}
\ket{L} = 
\begin{bmatrix}
1 \\
0 \\
0 \\
0
\end{bmatrix}
,
\qquad
\ket{R} = 
\begin{bmatrix}
0 \\
1 \\
0 \\
0
\end{bmatrix}
,
\qquad
\ket{D} = 
\begin{bmatrix}
0 \\
0 \\
1 \\
0
\end{bmatrix}
,
\qquad
\ket{U} = 
\begin{bmatrix}
0 \\
0 \\
0 \\
1
\end{bmatrix}
.
\end{align*}
The quantum walker on a plane, $\ZM^2$, can move to the left $\ket{L}$ or right $\ket{R}$ or down $\ket{D}$ or up $\ket{U}$ in each step. 

Put $\Theta_1=[0,1], \> \Theta_2=[0, 2 \pi]^3 \> (M_1=1, \> M_2=3)$ and $\theta_1 = \theta^{(1)}_1, \theta_2 = (\theta^{(1)}_2, \theta^{(2)}_2, \theta^{(3)}_2)$. Moreover, we define a one-parameter set of quantum coins introduced and intensively studied by Watabe et al. \cite{WatabeEtAl2008}:  
\begin{align}
U = U \left( \theta^{(1)}_1 \right) = 
\begin{bmatrix}
- \theta^{(1)}_1 & 1- \theta^{(1)}_1 & \sqrt{\theta^{(1)}_1 \left( 1- \theta^{(1)}_1 \right)} & \sqrt{\theta^{(1)}_1 \left(1- \theta^{(1)}_1 \right)} \\
1- \theta^{(1)}_1 & - \theta^{(1)}_1 & \sqrt{\theta^{(1)}_1 \left(1- \theta^{(1)}_1 \right)} & \sqrt{\theta^{(1)}_1 \left(1- \theta^{(1)}_1 \right)} \\
\sqrt{\theta^{(1)}_1 \left(1- \theta^{(1)}_1 \right)} & \sqrt{\theta^{(1)}_1 \left(1- \theta^{(1)}_1 \right)} & -\left(1- \theta^{(1)}_1 \right) & \theta^{(1)}_1 \\
\sqrt{\theta^{(1)}_1 \left(1- \theta^{(1)}_1 \right)} & \sqrt{\theta^{(1)}_1 \left(1- \theta^{(1)}_1 \right)} & \theta^{(1)}_1 & -\left(1- \theta^{(1)}_1 \right) 
\end{bmatrix},
\label{funfair05}
\end{align}
where $\theta^{(1)}_1 \in \Theta_1=[0, 1]$. Note that the QW determined by the quantum coin parameter $\theta^{(1)}_1 = 1/2$ becomes the four-state Grover walk on $\ZM^2$. Furthermore, we define a three-parameter set of initial states: 
\begin{align}
\varphi &= \varphi \left( \theta^{(1)}_2, \theta^{(2)}_2, \theta^{(3)}_2 \right) =
\frac{1}{2}
\begin{bmatrix}
1 \\
\exp (i \> \theta^{(1)}_2) \\
\exp (i \> \theta^{(2)}_2) \\
\exp (i \> \theta^{(3)}_2)  
\end{bmatrix}
.
\label{funfair06}
\end{align}
Here $(\theta^{(1)}_2, \theta^{(2)}_2, \theta^{(3)}_2) \in \Theta_2 = [0, 2 \pi]^3.$ 

In this section, we present three examples defined by ``Eqs. \eqref{funfair01} and \eqref{funfair02}",  ``Eqs. \eqref{funfair03} and \eqref{funfair04}", and ``Eqs. \eqref{funfair05} and \eqref{funfair06}", respectively. Of course, there are other options as well.

\section{Time-Series Model via QW \label{new}}
In this section, we propose a new time-series model based on the QW. We define the time series as a vector $D_n = \{{\bf x}_0, {\bf x}_1, \> {\bf x}_2, \ldots, {\bf x}_n \}$, where each element ${\bf x}_t = (x_t^{(1)}, x_t^{(2)}, \ldots , x_t^{(d)}) \in \RM^d \> (t=0,1,2, \ldots, n)$ is an $\RM$-valued $d$-dimensional vector. Each one of the $d$ values corresponds to the input variable measured in the time series. For example, if $d=2$ case, then each element ${\bf x}_n =(x_n^{(1)},x_n^{(2)}) \in \RM^2$ denotes that $x_n^{(1)}$ is the price and $x_n^{(2)}$ of the volume of a cryptocurrency at time $n$, respectively. 

If $D_n$ is given, we want to estimate a value ${\bf x}_{n+1}$ at the next time $n+1$ in a framework of our QW whose quantum coin is determined by $U = U (\theta_1) = U (\theta^{(1)}_1, \theta^{(2)}_1, \ldots, \theta^{(M_1)}_1)$ and initial state is given by $\varphi = \varphi (\theta_2) = \varphi (\theta^{(1)}_2, \theta^{(2)}_2, \ldots, \theta^{(M_2)}_2)$ at the origin. To estimate ${\bf x}_{n+1}$, we introduce the following time-dependent evaluation function $V_n = V_n (\theta_1, \theta_2)$ as 
\begin{align*}
V_n =  V_n (\theta_1, \theta_2) = \sum_{t=0}^{n} \>\> \sum_{||x||_1 \le t} \> || {\bf x} - {\bf x}_t ||_2 ^2 \> \mu_t ({\bf x}),
\end{align*}
where $||{\bf x}||_p = (|x^{(1)}|^p + |x^{(2)}|^p + \cdots + |x^{(d)}|^p)^{1/p} \> (0<p<\infty)$ for ${\bf x} = (x^{(1)}, x^{(2)}, \ldots , x^{(d)}) \in \RM^d$ and $\mu_t ({\bf x})$ is the probability measure for the QW at position $x$ and at time $t$.

Here we assume that ${\bf x}_0$ is the zero vector ${\bf 0}$ without loss of generality. This assumption corresponds to the fact that our QW starts at the origin. The following is our new algorithm to estimate ${\bf x}_{n+1}$ from a given data $D_n = \{{\bf x}_0= {\bf 0}, {\bf x}_1, \ldots, {\bf x}_n \}$. 

\par
\
\par
Step 1. Find a $(\theta_{1,n}^{\ast}, \theta_{2,n}^{\ast})$ such that $(\theta_{1,n}^{\ast}, \theta_{2,n}^{\ast})$ attains the minimum of $V_n (\theta_1, \theta_2)$.
\par
\
\par
Step 2. For the $(\theta_{1,n}^{\ast}, \theta_{2,n}^{\ast})$ given in Step 1, we compute $E(X_{n+1})$, where $E(X_{n+1})$ is the expectation of the position of quantum walker $X_{t}$ at time $t=n+1$ . If $(\theta_{1,n}^{\ast}, \theta_{2,n}^{\ast})$ is uniquely determined, then $E(X_{n+1})$ is considered as our estimated vector ${\bf x}_{n+1}$. So we put ${\bf x}_{n+1}^{\ast} = E(X_{n+1})$. If we have some options of $(\theta_{1,n}^{\ast}, \theta_{2,n}^{\ast})$, then we put ${\bf x}_{n+1}^{\ast}$ as the average of $E(X_{n+1})$ over these choices. If $V_n =  V_n (\theta_1, \theta_2)$ is a constant for any $\theta_1, \theta_2$, then we put ${\bf x}_{n+1}^{\ast} = {\bf x}_{n}$.
\par
\
\par
Step 3. By repeating above procedures, Steps 1 and 2, we obtain a sequence of estimated vectors $\{{\bf x}_1^{\ast}, {\bf x}_2^{\ast}, \ldots, {\bf x}_n^{\ast}, \ldots \}$.
\par
\
\par
Therefore, for each given data $D_n = \{{\bf x}_0= {\bf 0}, {\bf x}_1, \> {\bf x}_2, \ldots, {\bf x}_n \}$, we can get an estimated vector ${\bf x}_{n+1}^{\ast}$ by using the evaluation function $V_n$ based on the QW, sequentially. The details are explained for a two-state one-dimensional case in the next section.

\section{Two-State One-Dimensional Case \label{1d2s}}
In the first half of this section, we briefly give a definition of the two-state QW on $\ZM$ via a path counting method, which is equivalent to the definition mentioned in Section \ref{def}. The QW is a quantum version of the classical random walk with an additional degree of freedom called chirality. The chirality takes values left and right, and it means the direction of the motion of the walker. At each time step, if the walker has the left chirality, it moves one step to the left, and if it has the right chirality, it moves one step to the right. In this paper, we put
\begin{eqnarray*}
\ket{L} = 
\left[
\begin{array}{cc}
1 \\
0  
\end{array}
\right],
\qquad
\ket{R} = 
\left[
\begin{array}{cc}
0 \\
1  
\end{array}
\right],
\end{eqnarray*}
where $L$ and $R$ refer to the left and right chirality state, respectively.  

For the general setting, the time evolution of the walk is determined by a $2 \times 2$ unitary matrix called quantum coin, $U$, where
\begin{align*}
U =
\left[
\begin{array}{cc}
a & b \\
c & d
\end{array}
\right],
\end{align*}
with $a, b, c, d \in \mathbb C$. The matrix $U$ rotates the chirality before the displacement, which defines the dynamics of the walk. To describe the evolution of our model, we divide the quantum coin $U$ into two matrices:
\begin{eqnarray*}
P =
\left[
\begin{array}{cc}
a & b \\
0 & 0 
\end{array}
\right], 
\quad
Q =
\left[
\begin{array}{cc}
0 & 0 \\
c & d 
\end{array}
\right]
\end{eqnarray*}
with $U = P + Q$. The important point is that $P$ (resp. $Q$) represents that the walker moves to the left (resp. right) at position $x$ at each time step. In the present paper, we take $\varphi = {}^T [\alpha, \beta]$ with $\alpha, \> \beta \in \CM$ and $|\alpha|^2 + |\beta|^2 = 1$ as the initial qubit state, where $T$ is the transpose operator.

Let $\Xi_{n} (l,m)$ denote the sum of all paths starting from the origin in the trajectory consisting of $l$ steps left and $m$ steps right at time $n$ with $n=l+m$. For example, 
\begin{align*}
\Xi_2 (1,1) &= PQ + QP, \\
\Xi_4 (2,2) &= P^2 Q^2 + Q^2 P^2 + P Q P Q + Q P Q P + P Q^2 P + Q P^2 Q. 
\end{align*}
The probability that our quantum walker is in position $x \> (\in \ZM)$ at time $n \> (\in \ZM_{\ge})$ starting from the origin with $\varphi = {}^T [\alpha, \beta]$ with $\alpha, \> \beta \in \CM$ and $|\alpha|^2 + |\beta|^2 = 1$ is defined by 
\begin{align*}
P (X_{n} =x) = || \Xi_{n}(l, m) \> \varphi ||^2,
\end{align*}
where $n=l+m$ and $x=-l+m$. Let $\mu_n (x) = P (X_{n} =x)$. So we have
\begin{align*}
\mu_0 (0)
&=1,
\\
\mu_1 (-1) 
&= || P \varphi ||^2, \quad \mu_1 (1) = || Q \varphi ||^2,
\\
\mu_2 (-2) 
&= || P^2 \varphi ||^2, \quad \mu_2 (0) = || (PQ+QP) \varphi ||^2, \quad \mu_2 (2) = || Q^2 \varphi ||^2.
\end{align*}

We define the probability amplitude of the QW in position $x$ at time $n$ by 
\begin{equation*}
\Psi_{n}(x)=
\left[
\begin{array}{cc}
\Psi_{n}^{L}(x) \\
\Psi_{n}^{R}(x)
\end{array}
\right].
\label{eqn:quantum1}
\end{equation*}
Then we see that
\begin{align*}
P (X_{n} =x) = || \Psi_{n}(x) ||^2 = |\Psi_{n}^{L}(x)|^2 + |\Psi_{n}^{R}(x)|^2.
\end{align*}

In this section, we focus on the following setting which is essentially equivalent to one given in Eqs. \eqref{funfair01} and \eqref{funfair02}:
\begin{align}
U = U (\theta) = 
\begin{bmatrix}
\cos \theta & \sin \theta \\
\sin \theta & - \cos \theta
\end{bmatrix},
\qquad 
\varphi = \varphi (\xi) =
\begin{bmatrix}
\cos \xi \\
i \> \sin \xi
\end{bmatrix}
\qquad (0 \le \theta, \xi \le \pi/2).
\label{pac01}
\end{align}
Note that $\theta_1 = \theta_1^{(1)} = \theta$ and $\theta_2 = \theta_2^{(1)} = \xi.$

Let $\mu_n = \mu_n (\theta, \xi)$ be a probability measure for this QW. So we have a sequence of probability measures $\{ \mu_0, \mu_1, \ldots, \mu_n, \ldots \}$. 

Assume that $D_n = \{x_0, x_1, \ldots, x_n \}$ is a set of $\RM$-valued time-series data until time $n$. If $D_n$ is given, we want to estimate a next time $x_{n+1}$ by using a framework of our QW with a pair of parameters $(\theta, \xi)$. To do so, we introduced the following time-dependent evaluation function $V_n = V_n (\theta, \xi)$ as 
\begin{align*}
V_n =  V_n (\theta, \xi) = \sum_{t=0}^{n} \sum_{x=-t}^t \> \left( x - x_t \right)^2 \> \mu_t (x).
\end{align*}
We should note that we can not define the joint distribution of the QW such as $P(X_{t} =x, \> X_{t+1} =y).$

Here we assume that $x_0=0$ which corresponds to the fact that our QW starts at the origin. The following was our new algorithm to estimate $x_{n+1}$ from a given data $\{x_0=0, x_1, \ldots, x_n\}$. 

\par
\
\par
Step 1. Find a $(\theta_n^{\ast}, \xi_n^{\ast})$ such that $(\theta_n^{\ast}, \xi_n^{\ast})$ attains the minimum of $V_n (\theta, \xi)$.
\par
\
\par
Step 2. For the $(\theta_n^{\ast}, \xi_n^{\ast})$ given in Step 1, we compute $E(X_{n+1})$. If $(\theta_{1,n}^{\ast}, \theta_{2,n}^{\ast})$ is uniquely determined, then $E(X_{n+1})$ is considered as our estimated vector $x_{n+1}$. So we put $x_{n+1}^{\ast} = E(X_{n+1})$. If we have some options of $(\theta_{1,n}^{\ast}, \theta_{2,n}^{\ast})$, then we put $x_{n+1}^{\ast}$ as the average of $E(X_{n+1})$ over these choices. If $V_n =  V_n (\theta_1, \theta_2)$ is a constant for any $\theta_1, \theta_2$, then we put $x_{n+1}^{\ast} = x_{n}$.
\par
\
\par
Step 3. By repeating above procedures, Steps 1 and 2, we obtain a sequence of estimated values $\{x_1^{\ast}, x_2^{\ast}, \ldots, x_n^{\ast}, \ldots \}$.
\par
\
\par
From now on, we consider Steps 1 and 2 for $n=0$ and $n=1$ in the setting given by Eq. \eqref{pac01}. 
\par
\
\par
First we treat $n=0$ case. We want to obtain $x_1^{\ast}$ from $\{x_0 = 0\}$. This case is trivial, i.e., $x_1^{\ast} = x_0 = 0$. In fact, we see that
\begin{align*}
V_0 =  V_0 (\theta, \xi) = \left( 0 - x_0 \right)^2 \> \mu_0 (0) =0.
\end{align*}
Thus $V_0 = V_0 (\theta, \xi) = 0$ for any $\theta, \xi$. Therefore we can not determine $(\theta_0^{\ast}, \xi_0^{\ast})$. Then we put $x_1^{\ast}=x_0=0$.

Next we deal with $n=1$ case. We want to obtain $x_2^{\ast}$ from $\{x_0 = 0, x_1\}$. 
\par
\
\par
Step 1. We begin with 
\begin{align*}
V_1 
&=  V_1 (\theta, \xi) = \sum_{t=0}^{1} \sum_{x=-t}^t \> \left( x - x_t \right)^2 \> \mu_t (x)
\\
&= V_0 + \sum_{x=-1}^1 \> \left( x - x_1 \right)^2 \> \mu_1 (x)
\\
&= \left( -1 - x_1 \right)^2 \> \mu_1 (-1) + \left( 0 - x_1 \right)^2 \> \mu_1 (0) + \left( 1 - x_1 \right)^2 \> \mu_1 (1)
\\
&= \left( x_1^2 + 2 x_1 + 1 \right) \> \mu_1 (-1) + \left( x_1^2 - 2 x_1 + 1 \right) \> \mu_1 (1),
\end{align*}
since $V_0=0$ and $\mu_1 (0)=0$. Thus we have
\begin{align}
V_1 = \left( x_1^2 + 2 x_1 + 1 \right) \> \mu_1 (-1) + \left( x_1^2 - 2 x_1 + 1 \right) \> \mu_1 (1).
\label{neo01}
\end{align}
From now on we compute $\mu_1 (-1)$ and $\mu_1 (1)$ as follows. 
\begin{align*}
P \varphi 
&= 
\begin{bmatrix}
\cos \theta & \sin \theta \\
0 & 0
\end{bmatrix}
\begin{bmatrix}
\cos \xi \\
i \> \sin \xi
\end{bmatrix}
=
\begin{bmatrix}
\cos \theta \cos \xi + i \> \sin \theta \sin \xi \\
0
\end{bmatrix}
,
\\ 
Q \varphi 
&= 
\begin{bmatrix}
0 & 0 \\
\sin \theta & - \cos \theta
\end{bmatrix}
\begin{bmatrix}
\cos \xi \\
i \> \sin \xi
\end{bmatrix}
=
\begin{bmatrix}
0 \\
\sin \theta \cos \xi - i \> \cos \theta \sin \xi 
\end{bmatrix}
.
\end{align*}
By using these, we get
\begin{align}
\mu_1 (-1) 
&= || P \varphi ||^2 = \cos^2 \theta \> \cos^2 \xi + \sin^2 \theta \> \sin^2 \xi, 
\label{neo02a}
\\
\mu_1 (1) &= || Q \varphi ||^2 = \sin^2 \theta \> \cos^2 \xi + \cos^2 \theta \> \sin^2 \xi.
\label{neo02b}
\end{align}
Note that 
\begin{align}
\mu_1 (-1) + \mu_1 (1)=1. 
\label{neo03}
\end{align}
From Eqs. \eqref{neo01}, \eqref{neo02a}, \eqref{neo02b}, and \eqref{neo03}, we have
\begin{align*}
V_1 
&= \left( x_1^2 + 2 x_1 + 1 \right) \> \mu_1 (-1) + \left( x_1^2 - 2 x_1 + 1 \right) \> \mu_1 (1)
\\
&= \left\{ \mu_1 (-1) + \mu_1 (1) \right\} \> x_1^2 + 2 \left\{ \mu_1 (-1) - \mu_1 (1) \right\} \> x_1 + \left\{ \mu_1 (-1) + \mu_1 (1) \right\} 
\\
&= x_1^2 + 2 \cos (2 \theta) \> \cos (2 \xi) \> x_1 + 1.
\end{align*}
So we get 
\begin{align}
V_1 = V_1 (\theta, \xi) = x_1^2 + 2 \cos (2 \theta) \> \cos (2 \xi) \> x_1 + 1.
\label{neo04}
\end{align}
Moreover, 
\begin{align}
\frac{\partial V_1}{\partial \theta} 
&= - 4 \sin (2 \theta) \> \cos (2 \xi) \> x_1, 
\label{neo05a}
\\
\frac{\partial V_1}{\partial \xi} 
&= - 4 \cos (2 \theta) \> \sin (2 \xi) \> x_1.
\label{neo05b}
\end{align}
Here we consider three cases (i) $x_1 >0$, (ii) $x_1 =0$, (iii) $x_1 <0$ as follows. \par
\
\par
Case (i) $x_1 >0$. By Eqs. \eqref{neo04}, \eqref{neo05a}, and \eqref{neo05b}, we consider four cases in the following way.
\begin{align*}
&
\text{If} \> \left( \theta_1, \xi_1 \right) = \left(0, 0 \right), \> \text{then} \> V_1 = V_1 \left( 0, 0 \right) = x_1^2 + 2 x_1 + 1,
\\
&
\text{If} \> \left( \theta_1, \xi_1 \right) = \left(0, \frac{\pi}{2} \right), \> \text{then} \> V_1 = V_1 \left( 0, \frac{\pi}{2}  \right) = x_1^2 - 2 x_1 + 1,
\\
&
\text{If} \> \left( \theta_1, \xi_1 \right) = \left(\frac{\pi}{2}, 0 \right), \> \text{then} \> V_1 = V_1 \left(\frac{\pi}{2}, 0 \right) = x_1^2 - 2 x_1 + 1,
\\
&
\text{If} \> \left( \theta_1, \xi_1 \right) = \left(\frac{\pi}{2}, \frac{\pi}{2} \right), \> \text{then} \> V_1 = V_1 \left( \frac{\pi}{2}, \frac{\pi}{2}  \right) = x_1^2 + 2 x_1 + 1.
\end{align*}
Note that $x_1^2 + 2 x_1 + 1 > x_1^2 - 2 x_1 + 1$, since $x_1 >0$. Therefore we see that 
\begin{align*}
\left( \theta_1^{\ast}, \xi_1^{\ast} \right) = \left( 0, \frac{\pi}{2} \right), \> \left(\frac{\pi}{2}, 0 \right)
\end{align*}
and 
\begin{align*}
V_1 = V_1 \left(\theta_1^{\ast}, \xi_1^{\ast} \right) = x_1^2 - 2 x_1 + 1.
\end{align*}
\par
\
\par
Case (ii) $x_1 =0$. In this case, we see that $V_1 = V_1 (\theta, \xi) = 1$ for any $\theta, \xi$. 
\par
\
\par
Case (iii) $x_1 <0$. As in the case of (i), we see that 
\begin{align*}
\left( \theta_1^{\ast}, \xi_1^{\ast} \right) = \left( 0,0 \right), \> \left(\frac{\pi}{2}, \frac{\pi}{2} \right)
\end{align*}
and 
\begin{align*}
V_1 = V_1 \left(\theta_1^{\ast}, \xi_1^{\ast} \right) = x_1^2 + 2 x_1 + 1.
\end{align*}
\par
\
\par
Step 2. We compute $E(X_2)$ for the QW with $(\theta, \xi) = (\theta_1^{\ast}, \xi_1^{\ast})$ as follows. We begin with
\begin{align*}
E(X_2)
&= \sum_{x=-2}^{2} \> x \> \mu_2 (x)
= (-2) \> \mu_2 (-2) + 2 \> \mu_2 (2)
\\
&= (-2) \> || P^2 \varphi ||^2 + 2 \> || Q^2 \varphi ||^2
\\
&= -2 \> \cos^2 \theta \> \cos (2 \theta) \> \cos ( 2 \xi).
\end{align*}
Thus we have 
\begin{align}
E(X_2) = -2 \> \cos^2 \theta \> \cos (2 \theta) \> \cos ( 2 \xi).
\label{neo10}
\end{align}
By using Eq. \eqref{neo10}, we consider our estimated value $x_2^{\ast}$ in the following way.
\par
\
\par
Case (i) $x_1 > 0$. Furthermore, we consider two cases:
\par
\
\par
(a) if $( \theta_1^{\ast}, \xi_1^{\ast}) = (0,\pi/2)$, then Eq. \eqref{neo10} gives
\begin{align}
E(X_2) = 2.
\label{neo11a}
\end{align}
\par
(b) if $( \theta_1^{\ast}, \xi_1^{\ast}) = (\pi/2,0)$, then Eq. \eqref{neo10} gives
\begin{align}
E(X_2) = 0.
\label{neo11b}
\end{align}
In this case, we can not determine $x_2^{\ast}$. So we assume that two outcomes, $( \theta_1^{\ast}, \xi_1^{\ast} ) = (0,\pi/2)$ and $( \theta_1^{\ast}, \xi_1^{\ast}) = (\pi/2,0)$, are equally likely. That is, each event is selected with probability 1/2. Thus Eqs. \eqref{neo11a} and \eqref{neo11b} imply 
\begin{align*}
x_2^{\ast}=  2 \times \frac{1}{2} + 0 \times \frac{1}{2} = 1.
\end{align*}
Finally we obtain $x_2^{\ast}=1$. 
\par
\
\par
Case (ii) $x_1 =0$. We can not determine $x_2^{\ast}$. Thus we put $x_2^{\ast} = x_1 (=0).$
\par
\
\par
Case (iii) $x_1 < 0$. As in the case of (i), we consider two cases:
\par
\
\par
(a) if $( \theta_1^{\ast}, \xi_1^{\ast}) = (0,0)$, then Eq. \eqref{neo10} gives
\begin{align}
E(X_2) = -2.
\label{neo12}
\end{align}
\par
(b) if $( \theta_1^{\ast}, \xi_1^{\ast}) = (\pi/2,\pi/2)$, then Eq. \eqref{neo10} gives
\begin{align}
E(X_2) = 0.
\label{neo13}
\end{align}
In this case also, we can not determine $x_2^{\ast}$. Thus we assume that two outcomes, $( \theta_1^{\ast}, \xi_1^{\ast} ) = (0,0)$ and $( \theta_1^{\ast}, \xi_1^{\ast}) = (\pi/2,\pi/2)$, are equally likely. From Eqs. \eqref{neo12} and \eqref{neo13}, we see  
\begin{align*}
x_2^{\ast}=  (-2) \times \frac{1}{2} + 0 \times \frac{1}{2} = -1.
\end{align*}
Therefore we get $x_2^{\ast}=-1$. To summarize $n=1$ case, we have the following Table 1.
\begin{center}
\begin{tabular}{|c|c|c|}
\hline
$x_0 $ & $x_1$ & $x_2^{\ast}$ \\
\hline
$x_0 =0$ & $x_1 >0$ & $x_2^{\ast}=1$ \\
\hline
$x_0 =0$ & $x_1 =0$ & $x_2^{\ast}=x_1=0$ \\
\hline
$x_0 =0$ & $x_1 <0$ & $x_2^{\ast}=-1$ \\
\hline
\end{tabular}
\end{center}
\begin{center}
Table 1
\end{center}

\par
\
\par
Step 3. By Steps 1 and 2, for a given data $\{x_0=0, x_1\}$, we obtained a sequence of estimated values $\{x_1^{\ast} = x_0 = 0, x_2^{\ast} \}$. To summarize this, we present the following Table 2.
\begin{center}
\begin{tabular}{|c|c|}
\hline
$x_1^{\ast}$ & $x_2^{\ast}$ \\
\hline
$0$ & $ 1 \>\>\>\> (x_1 > 0)$ \\
\hline
$0$ & $ 0 \>\>\>\> (x_1 = 0)$ \\
\hline
$0$ & $ -1 \> (x_1 < 0)$ \\
\hline
\end{tabular}
\end{center}
\begin{center}
Table 2
\end{center}
If we consider general $n (\ge 2)$ case, then the following expression of $E(X_n)$ given by Proposition 2 in Konno \cite{Konno2002} is useful to compute $E(X_n)$ for $n \ge 3$ and $\theta \in [0, \pi/2)$:
\begin{align*}
E(X_n) 
&=
- (\cos \theta) ^{2(n-1)}
\Biggl[ n \cos (2 \theta) 
\\
& + \sum_{k=1}^{\left[{n-1 \over 2}\right]}
\sum_{\gamma =1} ^{k} \sum_{\delta =1} ^{k}
\left(-{\sin^2 \theta \over \cos^2 \theta} \right)^{\gamma + \delta} 
{k-1 \choose \gamma- 1} 
{k-1 \choose \delta- 1} 
{n-k-1 \choose \gamma- 1} 
{n-k-1 \choose \delta- 1} \\
& 
\times {(n-2k)^{2} \over  \gamma \delta} \{ n \cos (2 \theta) + \gamma + \delta \} \Biggr] \> \cos ( 2 \xi),
\end{align*}
where $[x]$ is the greatest integer that is less than or equal to $x \in \RM$. For example, we have 
\begin{align*}
E(X_3) = - \left\{ \left( 3 \cos^4 \theta + \sin^4 \theta \right) \> \cos (2 \theta) + \sin^2 \theta \> \sin^2 (2 \theta) \right\} \> \cos ( 2 \xi).
\end{align*}
We should remark that $\theta=\pi/2$ is a trivial case.

Furthermore, in order to determine $x_n^{\ast}$ for large $n$, it would be better to use a numerical method compared with the analytical method discussed as in the second half of this section \cite{KonnoEtAl2018}.



\section{Conclusions and Future Work \label{con}}
In this paper, we have proposed a new time-series method based on the QW. Additionally, we have discussed the method for a two-state one-dimensional case. On of the interesting problems would be to compare our method with the ARMA and/or GARCH models for some real data \cite{KonnoEtAl2018}, since our model does not impose a stationarity, but so do the ARMA and GARCH models.

Our model presented here is based on the QW. Recently the author extended the QW to a new walk called {\it quaternionic quantum walk} (QQW) determined by a unitary matrix whose component is quaternion \cite{Konno2015}. In general, the behavior of QQW is different from usual QW \cite{KonnoEtAl2016, Saito2017}, it is interesting to compare our method with a time-series one based on the QQW model. Moreover an extension from the QQW time-series model to the Clifford algebra time-series one would be also attractive.

\par
\
\par\noindent
{\bf Acknowledgment.} The author would like to thank Hiwon Yoon, Song-Ju Kim, Shinya Kawata, Masato Takei, Takashi Komatsu for useful discussions.
\par
\
\par

\begin{small}
\bibliographystyle{jplain}

\begin{thebibliography}{99}


\bibitem{AharonovEtAl1993}
Aharonov, Y., Davidovich, L., Zagury, N.:   
Quantum random walks. 
Phys. Rev. A {\bf 48}, 1687--1690 (1993)


\bibitem{AmbainisEtAl2001} 
Ambainis, A., Bach, E., Nayak, A., Vishwanath, A., Watrous, J.:
One-dimensional quantum walks. In: Proceedings of the 33rd Annual ACM Symposium on Theory of Computing, pp.37--49, 2001.


\bibitem{Bollerslev1986}
Bollerslev, T.: 
Generalized autoregressive conditional heteroskedasticity. 
Journal of Ecomometrics {\bf 31}, 307--327 (1986)


\bibitem{BoxJenkins1976}
Box, G. E. P., Jenkins, G. M.: 
Time Series Analysis: Forecasting and Control. 
Holden-Day, CA 1970.


\bibitem{CanteroEtAl2012b} 
Cantero, M. J., Gr\"unbaum, F. A., Moral, L., Vel\'azquez, L.: 
The CGMV method for quantum walks. 
Quantum Inf. Process. {\bf 11}, 1149--1192 (2012)


\bibitem{Engle1982} 
Engle, R. F.: 
Autoregressive conditional heteroscedasticity with estimates of the variance of United Kingdom inflation. 
Econometrica {\bf 50}, 987--1007 (1982)


\bibitem{Gudder1988}
Gudder, S. P.: 
Qunatum Probability.  
Academic Press Inc. CA (1988) 


\bibitem{IMKY2013}
Ichihara, A., Matsuoka, L., Kurosaki, Y., Yokoyama, K.: 
An analytic formula for describing the transient rotational dynamics of diatomic molecules in an optical frequency comb. 
Chin. J. Phys. {\bf 51}, 1230--1240 (2013)


\bibitem{Kitagawa2012}
Kitagawa, T.: 
Topological phenomena in quantum walks: elementary introduction to the physics of topological phases. 
Quantum Inf. Process. {\bf 11}, 1107--1148 (2012)


\bibitem{Konno2002}
Konno, N.: 
Quantum random walks in one dimension. 
Quantum Inf. Process. {\bf 1}, 345--354 (2002)


\bibitem{Konno2008b} 
Konno, N.: 
Quantum Walks. In: Quantum Potential Theory, Franz, U., and Sch\"urmann, M., Eds., Lecture Notes in Mathematics: Vol. 1954, pp. 309--452, Springer-Verlag, Heidelberg (2008)


\bibitem{Konno2013}
Konno, N.: 
A note on It\^o's formula for discrete-time quantum walk. 
J. Compu. Theo. Nanosci. {\bf 10}, 1579--1582 (2013)


\bibitem{Konno2015}
Konno, N.: 
Quaternionic quantum walks. 
Quantum Stud.: Math. Found. {\bf 2}, 63--76 (2015)


\bibitem{KonnoEtAl2018}
Konno, N., Kawata, S., Kim, S-J, Yoon, H.: In preparation.


\bibitem{KonnoEtAl2016}
Konno, N., Mitsuhashi, H., Sato, I.: 
The discrete-time quaternionic quantum walk on a graph. 
Quantum Inf. Process. {\bf 15}, 651--673 (2016) 


\bibitem{MW2013}
Manouchehri, K., Wang, J.: 
Physical Implementation of Quantum Walks. 
Springer, Berlin (2013) 


\bibitem{MKHY2011} 
Matsuoka, L., Kasajima, T., Hashimoto, M., Yokoyama, K.: 
Numerical study on quantum walks implemented on cascade rotational transitions in a diatomic molecule.
J. Korean Phys. Soc. {\bf 59}, 2897--2900 (2011)


\bibitem{MY2013}
Matsuoka, L., Yokoyama, K.: 
Physical implementation of quantum cellular automaton in a diatomic molecule. 
J. Compu. Theo. Nanosci. {\bf 10}, 1617--1620 (2013)


\bibitem{Meyer1996} 
Meyer, D. A.: 
From quantum cellular automata to quantum lattice gases. 
J. Statist. Phys. {\bf 85}, 551--574 (1996)


\bibitem{Nakamoto2008}
Nakamoto, S.:  
Bitcoin: A peer-to-peer electronic cash system.
https://bitcoin.org/bitcoin.pdf (2008)


\bibitem{OK2011}
Obuse, H., Kawakami, N.:
Topological phases and delocalization of quantum walks in random environments.
Phys. Rev. B {\bf 84}, 195139 (2011)


\bibitem{OkaEtAl2005}
Oka, T., Konno, N., Arita, R., Aoki, H.: 
Breakdown of an electric-field driven system: a mapping to a quantum walk.  
Phys. Rev. Lett. {\bf 94}, 100602 (2005)


\bibitem{P2013}
Portugal, R.: 
Quantum Walks and Search Algorithms. 
Springer, Berlin (2013) 


\bibitem{Saito2017}
Saito, K.: 
Probability distributions of quaternionic quantum walks. 
arXiv:1710.01482 (2017)


\bibitem{StefanakEtAl2012}
Stefanak, M., Bezdekova, I., Jex, I.: 
Continuous deformations of the Grover walk preserving localization. 
Eur. Phys. J. D {\bf 22}, 142 (2012)


\bibitem{VAndraca2008} 
Venegas-Andraca, S. E.: 
Quantum Walks for Computer Scientists. 
Morgan and Claypool (2008)


\bibitem{Venegas2013} 
Venegas-Andraca, S. E.: 
Quantum walks: a comprehensive review.  
Quantum Inf. Process. {\bf 11}, 1015--1106 (2012) 


\bibitem{WatabeEtAl2008}
Watabe, K., Kobayashi, N., Katori, M., Konno, N.: 
Limit distributions of two-dimensional quantum walks.
Phys. Rev. A {\bf 77} 062331 (2008)



\end{thebibliography}

\end{small}

\end{document}